\documentclass[aps,prl,reprint,groupedaddress]{revtex4-1}
\pdfoutput=1

\usepackage{amsmath}
\usepackage[utf8]{inputenc}
\usepackage{multirow}
\usepackage{graphicx}

\usepackage[colorlinks]{hyperref}
\hypersetup{allcolors=[rgb]{0,0,1}}

\renewcommand{\sec}[1]{\emph{#1.}---}

\begin{document}

\title{Invisible Neutrino Decay Resolves IceCube's Track and Cascade Tension}

\author{Peter B.~Denton}
\email{denton@nbi.ku.dk}
\author{Irene Tamborra}
\email{tamborra@nbi.ku.dk}
\affiliation{Niels Bohr International Academy and DARK, Niels Bohr Institute, University of Copenhagen, Blegdamsvej 17, 2100, Copenhagen, Denmark}

\date{\today}

\begin{abstract}
The IceCube Neutrino Observatory detects high energy astrophysical neutrinos in two event topologies: tracks and cascades.
Since the flavor composition of each event topology differs, tracks and cascades can be used to test the neutrino properties and the mechanisms behind the neutrino production in astrophysical sources.
Assuming a conventional model for the neutrino production, the IceCube data sets related to the two channels are in $>3\sigma$ tension with each other.
Invisible neutrino decay with lifetime $\tau/m=10^2$ s/eV solves this tension. Noticeably, it leads to an improvement over the standard non-decay scenario of more than $3\sigma$ while remaining consistent with all other multi-messenger observations. In addition, our invisible neutrino decay model predicts a reduction of $59\%$ in the number of observed $\nu_\tau$ events which is consistent with the current observational deficit. 
\end{abstract}

\maketitle

\sec{Introduction}The IceCube Neutrino Observatory measures high energy astrophysical neutrinos with energies reaching up to few PeVs~\cite{Aartsen:2013bka,Aartsen:2013jdh,Aartsen:2016xlq}. While numerous source candidates have been proposed to interpret the observed data, no clear picture has yet emerged~\cite{Anchordoqui:2013dnh,Meszaros:2015krr,Waxman:2015ues,Murase:2015ndr}. 

According to the conventional framework, adopted in this work, high energy astrophysical neutrinos are produced primarily by charged pion decay.
Charged pions decay to a muon and a muon neutrino, and the muon in turn decays on to a positron, electron neutrino, and a muon antineutrino resulting in a neutrino flavor ratio at the source of $\nu_e:\nu_\mu:\nu_\tau = 1:2:0$, each with approximately the same energy.
After neutrino oscillations, the flavor ratio at the Earth is roughly $1:1:1$ leading to the expectation that the spectral distributions of neutrinos will be the same for any flavor, see e.g.~\cite{Anchordoqui:2013dnh,Farzan:2008eg}.
This is independent of the source class since any mechanism that produces high energy neutrinos will do so dominantly as a result of charged pion decays. Hence, within this picture, the only possible result is equal fluxes for each flavor.

Single power law (SPL) and broken power law (BPL) fits have been considered to interpret the neutrino data~\cite{Aartsen:2015knd,Chen:2014gxa,Palladino:2016zoe,Anchordoqui:2016ewn,Palladino:2018evm,Chianese:2017jfa,Vincent:2016nut,Palomares-Ruiz:2015mka,Sui:2018bbh}. They favor a SPL, with a possible break to explain the excess of events below 100 TeV~\cite{Chianese:2017jfa,Denton:2018tdj}.

\begin{figure}
\centering
\includegraphics[width=\columnwidth]{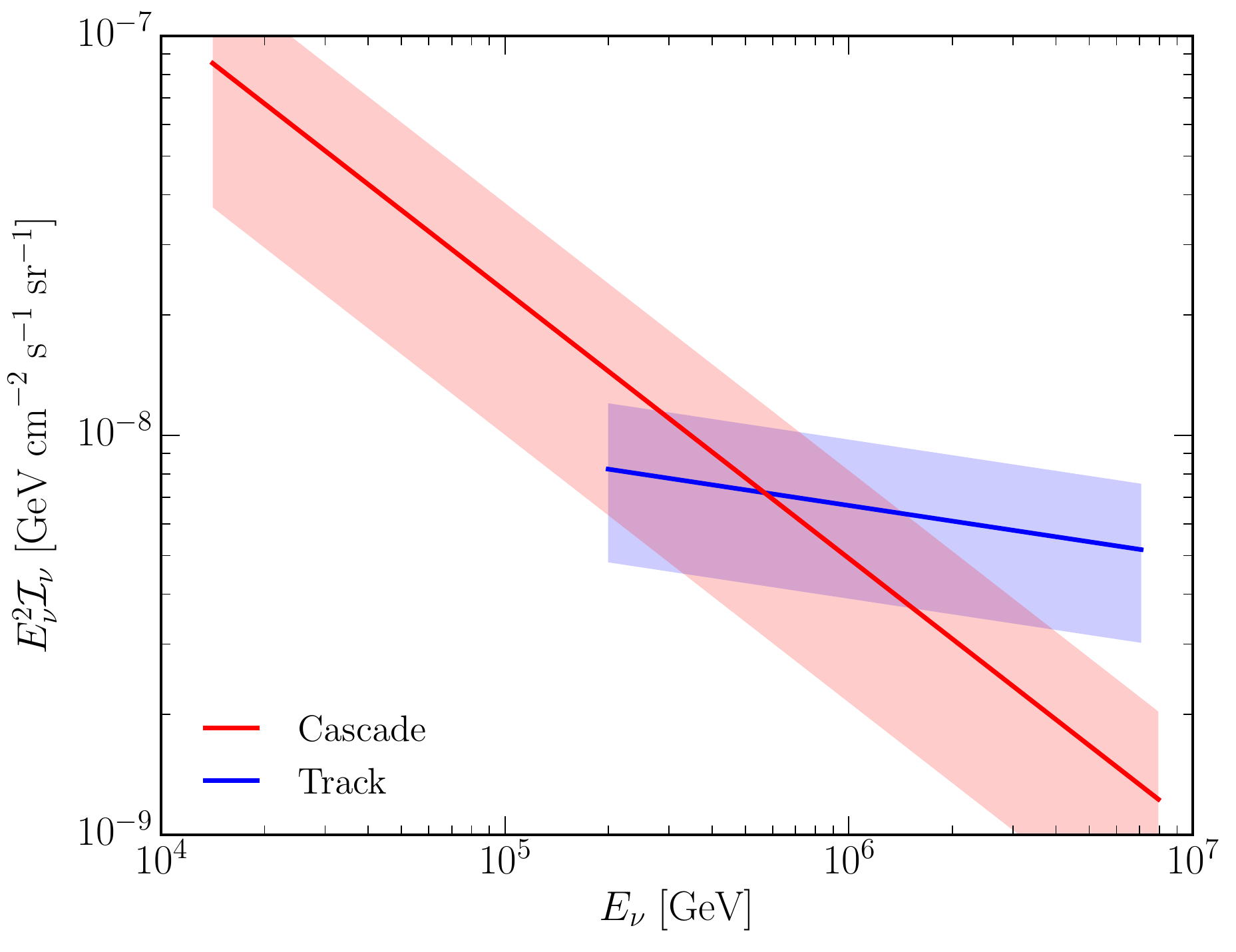}
\caption{IceCube track~\cite{Aartsen:2016xlq} and cascade~\cite{Niederhausen:2015svt} data samples. The tension between the two data samples is driven on the high energy end by the observation of six tracks with energies $E_\nu>1$ PeV.
On the low energy side there is an apparent excess of events in the cascade channel \cite{Denton:2018tdj}.}
\label{fig:ICdata}
\end{figure}

IceCube is partially sensitive to the flavor state of the neutrino through two distinct event topologies: track events resulting dominantly from $\nu_\mu$ interactions~\cite{Aartsen:2016xlq}, and nearly spherical cascade events resulting dominantly from $\nu_e$ and $\nu_\tau$ interactions~\cite{Niederhausen:2015svt}.
The IceCube Collaboration has interpreted each of these data sets in terms of the true per-flavor neutrino flux at the Earth under the assumption that the flavor ratio remains constant at $1:1:1$ for all energies and that the flux follows a SPL~\cite{Aartsen:2015knd}. It is found that those two different channels produce results in tension with each other~\cite{Aartsen:2016xlq}, as shown in Fig.~\ref{fig:ICdata}. 

IceCube finds that the best fit per-flavor astrophysical spectral index and normalization from the track analysis over $E_\nu \in[194$ TeV, $7.8$ PeV$]$ is $\gamma_{t,{\rm IC}}=2.13\pm0.13$, $\Phi_{t,{\rm IC}}=0.90^{+0.30}_{-0.27}$ \cite{Aartsen:2016xlq} and the best fit from the cascade analysis over $E_\nu \in[13$ TeV, $7.9$ PeV$]$ is $\gamma_{c,{\rm IC}}=2.67^{+0.12}_{-0.13}$, $\Phi_{c,{\rm IC}}=2.3^{+0.7}_{-0.6}$ \cite{Niederhausen:2015svt} where $\gamma_i$ is the spectral index and $\Phi_i$ is the flux normalization at $E_\nu = 100$ TeV in units of $10^{-18}$ GeV$^{-1}$ cm$^{-2}$ sr$^{-1}$ s$^{-1}$.

In this Letter, we combine spectral and flavor information simultaneously 
to investigate the tension between the data sets associated to the two event topologies. We explore several modifications to the standard picture of the high energy astrophysical neutrino flux beyond what is foreseen within the Standard Model \footnote{Neutrino oscillations already provide evidence of physics beyond the Standard Model in that they have mass. In this work, New Physics refers to physics beyond both the Standard Model and the fact that neutrinos have mass.}.

We determine the diffuse intensity at the Earth after oscillations, convert this into the per-flavor intensity from each of the track and cascade channels, and fit a power law to each assuming a $1:1:1$ flavor ratio to compare a model to IceCube's observations.
We then compare the normalizations and spectral indices to the measured ones by combining both tracks and cascades under the assumption that the correlation between the normalizations and spectral indexes are small.
Invisible neutrino decay provides a good fit to the data and is preferred over the Standard Model at more than $3\sigma$ removing the tension. Our proposed solution is not in contradiction with existing multi-messenger constraints and also explains the current deficit in the observation of $\nu_\tau$ events. 

\sec{Standard Neutrino Source Model}
For the sake of generality, we model the neutrino spectral distribution in such a way to be agnostic about the mechanism of the neutrino production, i.e.~$p\gamma$ or $pp$ interactions. We consider a general BPL model at the source parameterized by the break energy in the source frame $\tilde E_{\nu,b}$ and the change in the spectral index $\Delta$, such that the spectral index below the break energy is $\gamma$ and it is $\gamma+\Delta$ above it \cite{Anchordoqui:2013dnh,Meszaros:2015krr,Waxman:2015ues,Murase:2015ndr,Meszaros:2001vi}. The SPL case is then recovered for $\Delta = 0$.

This model is further generalized to the case where the break energy for neutrinos coming from muon decay ($\nu_e$ and $\nu_\mu$) is different than that from pion decay ($\nu_\mu$).
Pions and muons lose energy in $p\gamma$ sources, e.g.~in the presence of magnetic fields due to synchrotron losses and they may have separate break energies, $\tilde E_{\nu,b,\mu}$ and $\tilde E_{\nu,b,\pi}$.
For example, for synchrotron losses, the neutrino break energy scales like $m_i^{5/2}\tau_i^{-1/2}$ for $i\in\{\pi,\mu\}$ where $m$ ($\tau$) is the mass (lifetime) of the particle, so the ratio of the neutrino break energies is $R_{\pi,\mu} \equiv \tilde E_{\nu,b,\pi}/\tilde E_{\nu,b,\mu} \simeq 18.4$ when synchrotron cooling dominates.
The simpler BPL model introduced above is recovered when $R_{\pi,\mu}=1$.
Thus there are at most five free parameters in the BPL model: $\gamma$, $\Delta$, $\tilde E_{\nu,b}$, $R_{\pi,\mu}$, and the neutrino flux normalization $\Phi_\nu$. 

The IceCube neutrino flux is considered to be dominantly extragalactic and compatible with a diffuse origin~\cite{Denton:2017csz,Aartsen:2017ujz,Ando:2015bva,Aartsen:2015knd}. Hence, the expected diffuse neutrino intensity at the Earth for the flavor $\nu_\beta$ ($\beta=e,\mu,\tau$) is
\begin{equation}
\mathcal{I}_{\nu_\beta}=\sum_{\nu_\alpha} d_H\int_0^{z_{\max}}dz\frac{F_{\nu_\alpha}((1+z)E_\nu)\rho(z)}{h(z)}\bar P(\nu_\alpha \rightarrow \nu_\beta)\,,
\end{equation}
where $d_H=c/H_0$, $h(z)=\sqrt{(1+z)^3\Omega_m+\Omega_\Lambda}$, with $\Omega_m=0.308$, $\Omega_\Lambda=1-\Omega_m$, and $H_0=67.8$ km s$^{-1}$ Mpc$^{-1}$ \cite{Ade:2015xua}.
For the redshift evolution $\rho(z)$, we assume as benchmark case that the source luminosity density evolves as $(1+z)^\theta$ for $\theta=3$ up to a certain $z_c \simeq 1.5$ and it is constant for $z>z_c$~\cite{Gruppioni:2013jna}. Different redshift scalings for $\theta \in [0,4]$ and $z_c \in [0.5, 2]$ do not significantly affect our conclusions.
The averaged oscillation probability is $\bar P(\nu_\alpha \rightarrow \nu_\beta) = \sum_i|U_{\alpha i}|^2|U_{\beta i}|^2$ where $U$ is the standard mixing matrix \cite{Maki:1962mu,Pontecorvo:1967fh}.
For the mixing angles we take the latest global fit results~\cite{Esteban:2016qun,nu-fit:v3.2}.
The per-flavor flux from the source, $F_{\nu_\alpha}$, is either a SPL or a BPL.

We then compute the corresponding per-flavor intensity expected in the two event topologies; the track intensity roughly corresponds to the $\nu_\mu$ one, while the cascade one corresponds to the $\nu_e+\nu_\tau$ one (see the Appendix for technical details). 
A scan over all possible values of each model parameter is done to compare with the IceCube neutrino data through a $\chi^2$ test:
\begin{equation}
\chi^2=\sum_{i\in\{t,c\}}\left(\frac{\Phi_i-\Phi_{i,{\rm IC}}}{\sigma_{\Phi_{\nu,i}}}\right)^2+\left(\frac{\gamma_i-\gamma_{i,{\rm IC}}}{\sigma_{\gamma_i}}\right)^2\,;
\label{eq:chisq}
\end{equation}
where the sum runs on both neutrino event topologies, ($\Phi_i$, $\gamma_i$) are the normalization and spectral indices at the Earth which come from our calculations, and ($\Phi_{i,{\rm IC}}$, $\gamma_{i,{\rm IC}}$) fit the IceCube data.
For the SPL case with two free parameters ($\Phi_\nu$, $\gamma$) we find $\chi^2=13.4$ which corresponds to $3.23\sigma$ of tension.
When we expand the source model to the BPL case with four free parameters ($\gamma$, $\Delta$, $\tilde E_{\nu,b}$, $\Phi_\nu$) and $R_{\pi,\mu} = 1$, we find that the $\chi^2$ does not improve which results in $>3.66\sigma$ tension.
That is the BPL case is not preferred by the data with respect to the SPL. In addition, letting $R_{\pi,\mu}$ float freely only improves the fit to $\chi^2=10.7$ which is disfavored  at $>3.27\sigma$ and provides only marginal  improvement ($1.64\sigma$) over the BPL case.
In this case, the best fit point has $R_{\pi,\mu}>100$ and $\Delta$ large, similar to a  damped muon source.

Our findings confirm that adding a break to the source spectra provides marginal improvement to the data fit and that a SPL fit is justified. Most importantly, the standard neutrino source scenario is disfavored at $>3.2\sigma$ by the IceCube track and cascade data (see the left columns of Table \ref{tab:chisq} for a summary).
While muon cooling does provide both an energy and flavor dependent effect, it is not enough to resolve the tension due to the large mixing angles.
We expect that any mechanism which increases the relative number of $\nu_\mu$'s at the source (such as muon damping from synchrotron cooling) at high energy will equally increase the relative number of $\nu_\tau$'s after oscillations since $\theta_{23}\sim45^\circ$ is minimizing the effect.

\sec{Invisible Neutrino Decay}To solve the tension between the fits provided by the two event topologies, 
an interesting model modifying the flavor ratio in an energy dependent fashion during propagation is neutrino decay \cite{Beacom:2002vi,Shoemaker:2015qul,Moss:2017pur,Choubey:2017dyu}.
The latter is described by a new interaction term: $\mathcal L\supset g_{ij}\nu_i\nu_j\phi$ where $\phi$ is a new light ($m_\phi\lesssim m_\nu$) or massless scalar known as the Majoron, which could provide neutrinos with their masses \cite{Acker:1991ej, Gelmini:1980re,Chikashige:1980ui}. Specifically, we here focus on the invisible decay scenario where the decay products are a Majoron and a right handed neutrino (left handed antineutrino)~\cite{Gelmini:1980re,Chikashige:1980ui};
another  model of invisible neutrino decay is to unparticles \cite{Georgi:2007ek,Zhou:2007zq}.
Depending on the mass ordering and absolute mass scale, the decay products of visible neutrino decay may have significantly less energy.
For a steeply falling spectrum ($\gamma\gtrsim2$), visible decay becomes effectively invisible.

We assume that $\nu_1$ is stable since it has the least $\nu_\mu$ fraction since this can suppress the $\nu_\mu$ fraction at low energies.
This may be the case if the mass ordering is normal, as is currently favored at $2-3.4\sigma$~\cite{Esteban:2016qun,nu-fit:v3.2,deSalas:2017kay,globalfit,Capozzi:2018ubv}, and the Majoron has a mass between $\nu_1$ and $\nu_2$, or if $\nu_1$ is massless (or very light) and has no (significant) coupling to the Majoron.

The oscillation averaged probability is
\begin{equation}
\bar P(\nu_\alpha\to\nu_\beta)=\sum_{i=1}^3|U_{\alpha i}|^2|U_{\beta i}|^2e^{-\Lambda_i}\,,
\end{equation}
where $\Lambda_i\equiv d_H f(z)m_i/E_\nu\tau_i$ and $f(z)=\int_0^zdz'(1+z')^{-2}h^{-1}(z')$ is the corrected cosmological distance scaling for neutrino decay \cite{Baerwald:2012kc}.
We take $\Lambda_1=0$ and $\Lambda_2=\Lambda_3$; $\tau/m$, identical for $\nu_2$ and $\nu_3$, is our free parameter.

\begin{figure}
\centering
\includegraphics[width=\columnwidth]{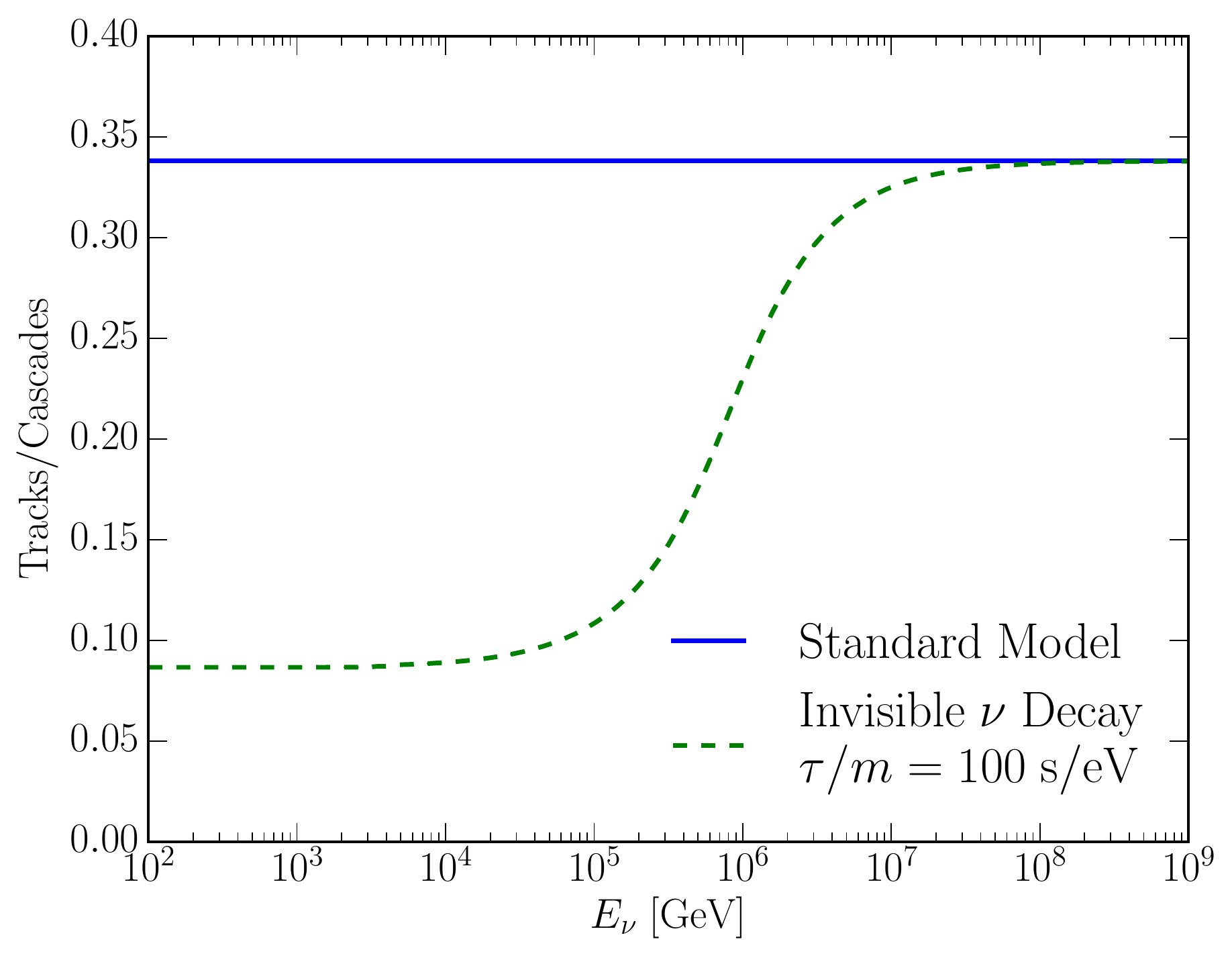}
\caption{The track to cascade ratio as a function of the neutrino energy. The invisible neutrino decay of $\nu_2$ and $\nu_3$ reduces the track and cascade ratio below 1 PeV up to $75\%$ with respect to the case where all neutrinos are stable. The deviation from the expected value of 0.5 for the standard case is mostly due to track misidentification wherein track events are sometimes misidentified as cascades (see the Appendix).}
\label{fig:nudecay}
\end{figure}
Figure~\ref{fig:nudecay} shows the modification of the track vs.~cascade ratio due to invisible neutrino decay within the model introduced above. One can check that in order to have an effect ($\Lambda_2,\Lambda_3\sim1$) within the region of interest of IceCube, we should have $\tau/m \sim10^2$ s/eV. 

\begin{figure}
\centering
\includegraphics[width=\columnwidth]{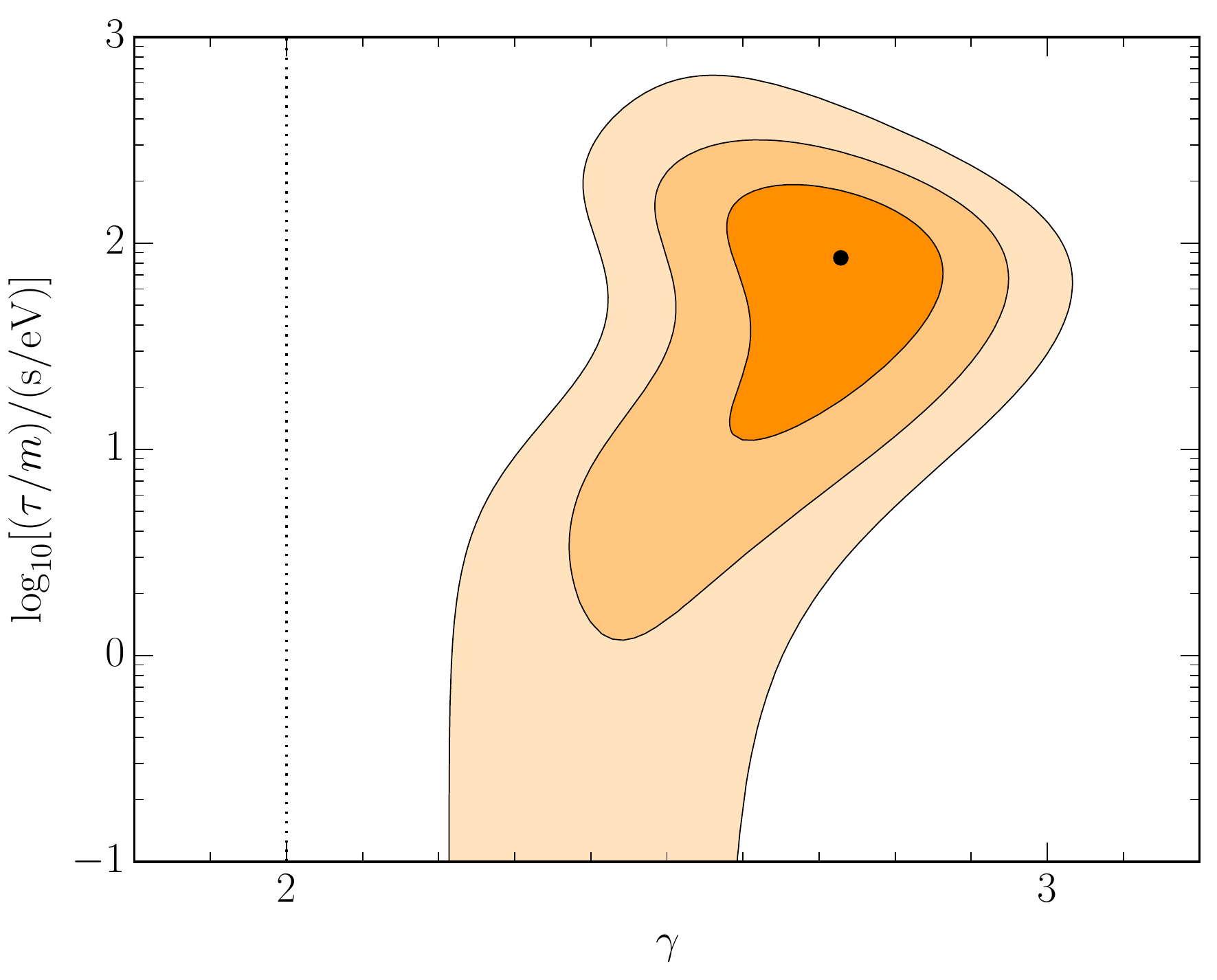}
\caption{The 2D $\chi^2$ projection for neutrino decay with a single power law astrophysical flux.
The shaded regions represent $1,2,3$ $\sigma$ for 2 d.o.f.
The best fit point of $\gamma=2.73$ and $\log_{10}[(\tau/m)/($s/eV$)]=1.93$, indicated with the dot, has $\chi^2=1.57$.
This includes a marginalization over the source normalization.
The slight preference for the full decay case over the $\nu$SM is because it modifies the relative normalization of the track and cascade diffuse intensities.}
\label{fig:Triangle}
\end{figure}

Minimizing the $\chi^2$ in the SPL only case with neutrino decay, we find $\chi^2=1.57$ with $\log_{10}[({\tau/m})/({\rm s/eV})]=1.93^{+0.26}_{-0.40}$.
At 1 d.o.f.~this represents a good fit, consistent with the data at $1.25\sigma$. It is an improvement over the stable neutrino case of $\Delta\chi^2=11.8$ showing that the neutrino decay scenario is preferred by the data over the standard stable neutrino case by $3.4 \sigma$.
The 2D $\chi^2$ projection of the source spectral index $\gamma$ and the neutrino lifetime $\tau/m$ is shown in Fig.~\ref{fig:Triangle}.
We note that $\tau/m$ is fairly well determined since it must give observable consequences within IceCube's region of interest.
Varying the redshift evolution power $\theta$ produces a fairly small effect with the best fit value of $\tau/m$ and the $\chi^2$ slightly changes  with $\tau/m$ increasing with $\theta$.
If we extend our fit to the BPL source model, the best fit point does not change at all and $\Delta=0$ is preferred, see 
Table \ref{tab:chisq}~\footnote{A newer unpublished analysis from the IceCube Collaboration~\cite{Aartsen:2017mau} slightly changes the various qualities of fit related to the track and cascade datasets. Given the different energy ranges, the tension between the track and cascade data decreases to $2.5 \sigma$ for the SPL. However, we find that this does not significantly change our conclusions and neutrino decay is still preferred at $2.8 \sigma$. This trend has also been confirmed from the preliminary results presented at Neutrino 2018 \cite{taboada_ignacio_2018_1286919}.}.

Our findings should be compared with existing bounds on invisible neutrino decay. The best terrestrial constraints on invisible $\nu_3$ decay come from atmospheric and long-baseline data: $\log_{10}[({\tau_3/m_3})/({\rm s/eV})]>-9.52$~\cite{GonzalezGarcia:2008ru,Pagliaroli:2016zab}; the best terrestrial constraints on invisible $\nu_2$ decay are from solar neutrinos and are $\log_{10}[(\tau_2/m_2)/({\rm s/eV})]>-3.15$~\cite{Berryman:2014qha,Picoreti:2015ika}.
Hints for $\nu_3$ invisible decay exist at $\log_{10}[(\tau_3/m_3)/({\rm s/eV})]\sim-11$~\cite{Gomes:2014yua,Choubey:2018cfz}.
\begin{table}[h]
\renewcommand{\arraystretch}{1.1}
\centering
\caption{The $\chi^2$ and significance for the single power law (SPL) and broken power law (BPL) models, along with the best fit source spectral index and neutrino lifetime.
Here we fix $R_{\pi,\mu}=1$ for the BPL model, see text.
The BPL models have as many or more parameters than data points; only a lower limit on the significance can be placed by taking 1 d.o.f.}
\label{tab:chisq}
\begin{tabular}{c|c|c|c|c}
\multirow{2}{*}{Model}&\multicolumn{2}{c|}{Standard Model}&\multicolumn{2}{c}{Invisible $ \nu$ Decay}\\\cline{2-5}
&SPL&BPL&SPL&BPL\\\hline
$\chi^2$&13.4&13.4&1.57&1.57\\
$\sigma$&$3.23$&$>3.65$&$1.25$&$>1.25$\\
$\gamma$&$2.4\pm0.10$&-&$2.73\pm0.10$&-\\
$\log_{10}(\frac{\tau/m}{\rm s/eV})$&-&-&$1.93^{+0.26}_{-0.40}$&$1.93^{+0.26}_{-0.40}$
\end{tabular}
\end{table}

Strong constraints, in apparent contradiction with our findings, have been derived from SN 1987A: $\log_{10}[({\tau/m})/({\rm s/eV})]\gtrsim 5$~\cite{Hirata:1987hu}; however these constraints only apply to $\bar{\nu}_e$ measurements under the assumption that all neutrino mass eigenstates are decaying and should be considered with caution. 
Even in the case of full $\nu_2$ and $\nu_3$ decay, the $\bar{\nu}_e\to\bar{\nu}_e$ oscillation averaged probability would be suppressed by 16\% which is still smaller than the SN 1987A statistical uncertainties ($\sim20\%$) and current theoretical uncertainties.
IceCube data has been used to place a constraint on the neutrino lifetime at $\log_{10}[({\tau/m})/({\rm s/eV})] \gtrsim1$ by assuming that neutrinos do not fully decay within the IceCube energy range \cite{Pagliaroli:2015rca,Bustamante:2016ciw} which is not the case considered here.
The most stringent constraints on the lifetime of neutrinos have been derived from cosmic microwave background data at the level of $\log_{10}[({\tau/m})/({\rm s/eV})] \gtrsim 11$ \cite{Hannestad:2005ex}. Noticeably, these bounds can be alleviated in the event that only one or two neutrinos decay and the remaining ones are free streaming~\cite{Bell:2005dr,Archidiacono:2014nda,Gariazzo:2014pja} and are therefore not in contradiction with our findings.
Interestingly, neutrino decay with parameters similar to our model was  proposed as an alternate solution to the solar neutrino problem \cite{Bahcall:1986gq}.

\sec{Other Possible Interpretations}Another possible explanation of the tension between the track and cascade data sets is the decay of dark matter (DM) \cite{Aartsen:2018mxl} to electron neutrinos ($\chi \rightarrow \nu_e \bar{\nu}_e$).
We focus on DM decay instead of annihilation as the galactic anisotropy constraints \cite{Ahlers:2015moa,Denton:2017csz,Aartsen:2017ujz} are weaker for DM decay since the DM annihilation more peaked towards the galactic center.
In order to estimate the expected track and cascade distribution, the galactic and extragalactic diffuse intensity of neutrinos is computed, including electroweak corrections, by using {\sc Pythia} 8.2 \cite{Sjostrand:2014zea} and a Navarro-Frenk-White galactic DM profile \cite{Navarro:1995iw}.

While a good quality of fit ($\chi^2<1$) is found in a SPL+DM model with 4 parameters ($\tau_\chi$ s, $m_\chi$ TeV, $\Phi_\nu$, $\gamma$), this model has a number of undesirable properties.
The galactic contribution to the flux peaks at energies below the cascade flux sensitivity and contributes due to the typical energy uncertainty of cascades are $\gtrsim15\%$ \cite{Aartsen:2015zva};
this results in a contribution to the cascade flux at low energies due to the energy uncertainty, but a minimal contribution to the track flux (after oscillations).
The resultant  peak flux is larger than the measured flux at energies just below the region of interest for IceCube's cascade analysis.
From SU(2) symmetry there will be an $e^+e^-$ channel,  leading to $\gamma$-rays from electroweak corrections constrained by {\it Fermi}-LAT \cite{Murase:2015gea}.
Finally, this fit requires a short DM lifetime which is strongly constrained by the cosmic microwave background and bounds from the reionization epoch,  the best fit values being $\tau_\chi\sim10^{23}$ s, $m_\chi\sim10$ TeV~\cite{Liu:2016cnk,Slatyer:2016qyl}.
All considered,  DM decay does not seem to resolve this tension.

Several additional effects could provide an energy and flavor dependent modification of the standard neutrino flux from an astrophysical source. For example, the Glashow resonance occurs when a $\bar\nu_e$ with $E_\nu=6.3$ PeV scatters off an electron in the ice creating an on-shell $W^-$ \cite{Glashow:1960zz} increasing IceCube's sensitivity in that energy range considerably.
IceCube performs their fits assuming that $\mathcal{I}_\nu=\mathcal{I}_{\bar\nu}$.
While this is generally the case if neutrinos are mainly produced through $pp$ interactions, it won't be the case if the main neutrino production channel is $p\gamma$ interactions \cite{Nunokawa:2016pop}.
For the SPL case $\mathcal{I}_{\nu_e}/\mathcal{I}_{\bar\nu_e}\simeq3.5$ which would somewhat harden the cascade spectrum, but would not be enough to reduce the tension of the fit.

Another option that could alleviate the track vs.~cascade fit tension is neutron decay sources.
Neutrons decay to $\bar\nu_e$'s and are produced alongside charged pions in $p\gamma$ interactions (as well as in $pp$ interactions) and are thus expected to provide an additional contribution of $\nu_e$'s to the high energy astrophysical neutrino flux.
The energy of neutrinos from neutrons is suppressed by about two orders of magnitude compared with those from pion decay.
However, for a spectral index $\gtrsim2$ as in our case, this contribution is subleading.
Neutrons also result from photodisintegration of heavy ions in dense sources, although this flux is also suppressed compared to the standard contribution by at least an order of magnitude \cite{Biehl:2017zlw,Rodrigues:2017fmu}.

In addition, non-standard neutrino interactions with ultralight mediators ($m_{Z'}\ll1$ eV) as well as pseudo-Dirac neutrino models~\cite{Wolfenstein:1981kw,Pakvasa:2012db} may also affect the track vs.~cascade ratio. 
However, in both cases, we expect an impact on the neutrino data set that is smaller than the one induced by the invisible neutrino decay scenario.

\sec{A Solution to the $\nu_\tau$ Observational Deficit}The IceCube detector is expected to observe 2 or 3 $\nu_\tau$ events in the energy range of interest~\cite{Aartsen:2017mau,Palladino:2018qgi}. However, currently no $\nu_\tau$ events are observed. The assumption of invisible neutrino decay for the $\nu_2$ and $\nu_3$ eigenstates would induce a reduction of $\mathcal{I}_{\nu_\tau}$ of $80\%$ below 1 PeV which convolved with the detection efficiency leads to a $\sim59\%$ reduction in the number of $\nu_\tau$ events for our best fit value $\tau/m = 10^2$~s/eV. The invisible neutrino decay could then also explain the current deficit of $\nu_\tau$ events.

\sec{Conclusions}The IceCube  Observatory detects high energy astrophysical neutrinos through two event topologies: tracks and cascades. By simultaneously taking advantage of the energy and flavor information present in the two data sets, for the first time we have placed strong constraints on the consistency of the data with the standard source picture.
A conventional model for the neutrino production in astrophysical sources is unable to simultaneously explain the track and cascade data at $>3\sigma$.

We tested several New Physics models and found that the invisible neutrino decay of $\nu_2$ and $\nu_3$ with $\tau/m=10^2$ s/eV is preferred by the IceCube data by $3.4\sigma$ and is consistent with all other existing constraints.
While this model is more natural in the normal mass ordering, it is consistent with either ordering.
In addition, a model of visible decay in $\nu_1$ may provide additional improvements to the fit by producing additional $\nu_1$'s (mostly $\nu_e$'s) at lower energies.
Interestingly, our model also predicts a $59\%$ reduction in the number of expected $\nu_\tau$ events reconciling the current observational deficit. 

As more neutrino data arrives with the advent of IceCube-Gen2 \cite{Aartsen:2014njl} and KM3NeT \cite{Adrian-Martinez:2016fdl} and the spectral distributions will be defined more precisely for both event topologies, it will be possible to further test our result.

\sec{Acknowledgments}We are grateful to Mauricio Bustamante, Steen Hannestad, Rebecca Leane, Orlando Peres, and Mohamed Rameez for useful discussions.
PBD and IT acknowledge support from the Villum Foundation (Project No.~13164) and the Danish National Research Foundation (DNRF91).
PBD thanks the Danish National Research Foundation (Grant No.~1041811001) for support.
The work of IT has also been supported by the Knud H\o jgaard Foundation and the Deutsche Forschungsgemeinschaft through Sonderforschungbereich SFB 1258 ``Neutrinos and Dark Matter in Astro- and Particle Physics'' (NDM).

\bibliography{Tracks_Cascades}

\section{Appendix{\label{sec:appendix}}}
\appendix
\sec{Per-flavor Flux Reconstruction}
Converting the flux observed at the Earth after oscillations and decay in a given channel to the per-flavor true flux includes corrections due to neutral current interactions, track misidentification in a given IceCube detection channel \cite{Aartsen:2015ivb}, and $\tau\to\mu+2\nu$ decays, each of which is accounted for in our analysis and contributes only a sub-leading effect on our results.
The details of these corrections are presented here.

The IceCube Collaboration reports the per-flavor flux from the track and cascade analyses.
The track analysis is dominantly the result of $\nu_\mu$ charged current (CC) interactions and the cascade analysis is dominantly the result of $\nu_e$ and $\nu_\tau$ CC interactions.

For clarity, the relevant flux and intensity terms are now defined again.
The flux of a single source of $\nu_\alpha$ is $F_{\nu_\alpha}$ and is normalized by $\Phi_\nu$ which is a free parameter in the fits.
The diffuse intensity at the Earth after oscillations and decay of $\nu_\alpha$ is $\mathcal I_{\nu_\alpha}$ (note that $F_{\nu_\alpha}$ and $\mathcal I_{\nu_\alpha}$ both refer to the sum of neutrinos and antineutrinos unless otherwise mentioned).
We take $f_{\rm mis}=0.3$ as the fraction of CC $\nu_\mu$ interactions that are misidentified as cascades~\cite{Aartsen:2015ivb} and $f_{\rm CC}=0.7$ as the fraction of neutrino events that undergoes a CC interaction, while the rest undergo a NC interaction depositing $\sim1/3$ of the energy in the detector \cite{Gandhi:1995tf}.
The branching ratio of $\tau\to\mu+2\nu$ is $f_{\tau\mu}=0.174$~\cite{Patrignani:2016xqp}.
Thus the track ($t$) intensity is related to the neutrino flux by
\begin{equation}
\mathcal{I}_t(E_\nu)=f_{\rm CC}(1-f_{\rm mis})\mathcal{I}_{\nu_\mu}(E_\nu)+f_{\rm CC}f_{\tau\mu}\mathcal{I}_{\nu_\tau}(3E_\nu)\,.
\label{eq:It}
\end{equation}
In order to convert this into the per-flavor flux, we follow IceCube's approach of assuming that $\mathcal{I}_{\nu_e}=\mathcal{I}_{\nu_\mu}=\mathcal{I}_{\nu_\tau}$ and that they are described by a SPL.
Then the per-flavor intensity at the Earth from the track data set is
\begin{equation}
\mathcal{I}_{t,{\rm pf}}(E_\nu)=\frac{\mathcal{I}_t(E_\nu)}{f_{\rm CC}(1-f_{\rm mis})+f_{\rm CC}f_{\tau\mu}3^{-\gamma_t}}\ ,
\label{eq:Itpf}
\end{equation}
where $\gamma_t$ is the result of a power law fit to $\mathcal{I}_t(E_\nu)$.

Similarly, the cascade ($c$) intensity is related to the neutrino flux at the Earth by
\begin{multline}
\mathcal{I}_c(E_\nu)=f_{\rm CC}[\mathcal{I}_{\nu_e}(E_\nu)+f_{\rm mis}\mathcal{I}_{\nu_\mu}(E_\nu)+(1-f_{\tau\mu})\mathcal{I}_{\nu_\tau}(E_\nu)]\\
+(1-f_{\rm CC})\sum_{\alpha\in\{e,\mu,\tau\}}\mathcal{I}_{\nu_\alpha}(3E_\nu)\,.
\label{eq:Ic}
\end{multline}
Then the per-flavor flux at the Earth from the cascade data set is
\begin{equation}
\mathcal{I}_{c,{\rm pf}}(E_\nu)=\frac{\mathcal{I}_c(E_\nu)}{f_{\rm CC}[1+f_{\rm mis}+(1-f_{\tau\mu})]+3(1-f_{\rm CC})3^{-\gamma_c}}\,.
\label{eq:Icpf}
\end{equation}

If $\mathcal{I}_{\nu_\alpha}(E_\nu)$ is a power law and the diffuse intensity of each flavor is the same then these definitions recover the correct true neutrino intensity, while also allowing for different spectra at the Earth in terms of both deviations from a SPL and different intensities for different flavors.
Compared with setting $\mathcal I_t \simeq \mathcal I_{\nu_\mu}$ and $\mathcal I_c \simeq \mathcal I_{\nu_e}+\mathcal I_{\nu_\tau}$, including these corrections is a $\lesssim1\%$ correction on the diffuse intensities.
Hence the diffuse intensities of each flavor are related to the track and cascade intensities $\mathcal I_i$ for $i\in\{$t$,$c$\}$ by Eqs.~\ref{eq:It} and \ref{eq:Ic}.
Finally, these are converted into the per-flavor intensities $\mathcal I_{i,{\rm pf}}$ under the assumption of a SPL and $\mathcal{I}_{\nu_e}=\mathcal{I}_{\nu_\mu}=\mathcal{I}_{\nu_\tau}$ flavor ratio by Eqs.~\ref{eq:Itpf} and \ref{eq:Icpf}; these have normalizations at 100 TeV of $\Phi_i$ which can then be compared with the data.

\end{document}